\documentclass[pra,twocolumn,amsmath,amssymb,superscriptaddress]{revtex4-1}

\usepackage{epsfig,amsmath}
\usepackage{subfigure}
\usepackage{graphicx}
\usepackage{dcolumn}
\usepackage{stmaryrd}
\usepackage{mathrsfs}
\usepackage{pifont}
\usepackage{amsthm}
\usepackage{amssymb}
\usepackage{bm}
\usepackage{latexsym}
\usepackage[colorlinks=true,linkcolor=blue,citecolor=blue]{hyperref}
\usepackage{color}
\usepackage{epstopdf}

\begin{document}

\title{Magnon-mediated quantum battery under systematic errors}

\author{Shi-fan Qi}
\affiliation{Department of Physics, Zhejiang University, Hangzhou 310027, Zhejiang, China}

\author{Jun Jing}
\email{Email address: jingjun@zju.edu.cn}
\affiliation{Department of Physics, Zhejiang University, Hangzhou 310027, Zhejiang, China}

\date{\today}

\begin{abstract}
  Quantum battery is one of the most prominent micro-devices in the rapid-developing field of quantum thermodynamics. We propose a long-range charging protocol in which both battery and charger are consisted of a many-spin system. The battery and charger are indirectly connected via a quantum wire supported by the Kittle mode of a magnon system. Counter-intuitively, we find that the undesired spin-spin couplings among both battery and charger can be used to promote the charging performance in comparison to the noninteracting condition. And a ``sweet'' spot of the average coupling strength allows the battery to store energy to the maximum possible value. Moreover, we apply the quantum-state-diffusion equation to estimate the robustness of our magnon-mediated charging protocol to the fluctuation in the magnon frequency.
\end{abstract}

\maketitle

\section{Introduction}

The figure of merit of a quantum battery (QB) is quantified in the capacity to fully transfer the stored energy to the consumption cells in a useful way, i.e., the skill of extracting useful work~\cite{work}, and the maximum average charging power~\cite{highpowerbattery}. Recently, a worldwide interest arises in pursuit of the high-capacity and high-power quantum batteries~\cite{entangbattery,powerbattery,chargebattery,disscharge,envirbattery,
twophotonbattery,openbattery,heatbattery,thermalcharge} to exceed their classical counterparts. It is thus significant to exploit the existing quantum models, configurations, platforms as well as control techniques to find optimal charging protocols, that are able to store and dispatch more energy within a shorter timescale in a stable way.

Aim to identify and demonstrate the so-called quantum advantage for QBs, many works focus on understanding how the presence of quantum coherence or entanglement can affect the energy storage in both single-body~\cite{chargebattery,drivenbattery,dissipative,floquet} and many-body~\cite{spinchain,highpowerbattery,sykbattery,powerfulharmonic} QBs. The first study~\cite{work} presented in 2013 demonstrates that the system entanglement leads to a speed-up in the work-extraction (energy-discharging) process. The parallel charging and the collective charging were identified as two distinct charging schemes~\cite{powerbattery,enhancepower} in terms of the battery-charger connection. Two implementations of collective charging~\cite{spinchain,highpowerbattery} were proposed later on to claim their advantage over the parallel charging schemes. Nevertheless, it was debated~\cite{classical,bound} that the preceding collective charging of the many-body QBs does not hold any genuine quantum advantage. Stability is also an important feature of quantum batteries~\cite{meabattery}, which is characterized by the energy-backflow suppression after the charging is completed. Stimulated Raman adiabatic passage was applied to ensure a stable adiabatic charging process~\cite{stablebattery}. A general approach based on the time-dependent Hamiltonian provided a treatment to the stability problem in the discharging process to an available consumption cell~\cite{chargeswitchable}.

Different prototypes of quantum batteries have been devised in potential laboratory platforms for quantum information and quantum optics. In hybrid quantum systems, one-dimensional Heisenberg XY chain can be used as a many-body quantum battery~\cite{spinchain,bound,manybodybattery,disorder} charged by an external magnetic field. The Dicke model~\cite{highpowerbattery,chargebattery,asymptotic,openbattery,powerfulharmonic}, that describes an array of $N$ noninteracting qubits coupled to a common cavity mode, has been widely considered as a proper quantum battery. The Dicke quantum battery charged by the cavity mode, which in most cases has to be prepared as a Fock state with a high excitation number, is however a challenge in experiment. Converting the charger from a continuous-variable system to a discrete-variable system, such as quantum spins, might overcome the challenge for providing sufficient energy to the battery. Yet the ordinary dipole-dipole interaction between the battery spin and the charger spin is a short-range force, which means the battery (spins) must be located nearby the charger (spins). Then a non-local medium connecting the battery and the charger, such as a global environment~\cite{envirbattery}, becomes necessary and has been proposed to realize a long-range charging. But such a protocol demands an environment in a deep non-Markovian regime.

Hybrid quantum systems mediated by magnon have become novel platforms for the promising quantum technologies~\cite{magnoncavity,magnon,magnonhybrid,magnons}. A long-range and high-fidelity state transfer can be available by virtue of the efficient coupling of magnonic excitations to nanometer-scale magnetic spin emitters and the magnon-mediated spin-spin interactions~\cite{spinmagnoncoupling}. In this work, we model both quantum battery and charger as many-spin systems, which are coupled to a magnon system serving as a quantum ``wire''. The overall system can be used to implement a long-range charging protocol, which can be switched on and off by tuning the magnon frequency via the bias magnetic field. Note magnons as bosonic quasiparticles describe collective excitations of spins in materials with a finite magnetic moment. The magnon system in our work is a single-crystal yttrium iron garnet (YIG) sphere, which is in a uniform magnetostatic mode, also called the Kittel mode, with the lowest order of excitation~\cite{kittelmode}. The frequency of the Kittel mode is given by $\omega_m=g^*\mu_BB_0/\hbar$, where $g^*$ is the g-factor, $\mu_B$ is the Bohr magneton, and $B_0$ is the amplitude of the external magnetic field~\cite{magnon,magnonqubit,magnonqubit2}. More interesting, it is found that if the indirect coupling among the spins (of either battery or charger) mediated by the magnon mode could neutralize their direct coupling from the undesired crosstalk, then the effective Hamiltonian for the overall system supports a perfect charging process. Though the crosstalk among battery cells were thought to be a negative factor in the energy storage~\cite{powerfulharmonic}.

The rest part of this work is structured as following. In Sec.~\ref{model}, we introduce the hybrid model for our quantum battery and derive an effective Hamiltonian between the charger and battery mediated by the magnon mode (the details can be found in appendix~\ref{appa}). That enables a general long-range charging protocol of such a quantum battery. Various charging scenarios, including one battery cell charged by one charger spin, one battery cell charged by many charger spins, and many battery cells charged by many charger spins, are investigated in Sec.~\ref{oneone}, Sec.~\ref{multione}, and Sec.~\ref{multimulti}, respectively. By tuning the direct coupling strengths among spins, we find an optimized charging could be realized at a sweet spot. In Sec.~\ref{qsdprocess}, we consider the effect of a systematic error over the magnon frequency on the charging performance by virtue of a non-Markovian quantum-state-diffusion (QSD) equation~\cite{qsd0,qsd,qsd1,qsd2,qsd3}. The exact solution shows that our magon-mediated protocol is robust to the error. Discussion and conclusion are provided in Sec.~\ref{conclusion}.

\section{MODEL}\label{model}

\begin{figure}[htbp]
\centering
\includegraphics[width=0.4\textwidth]{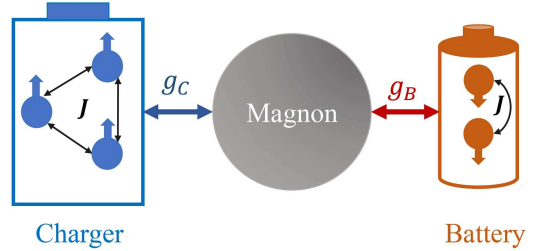}
\caption{(Color online) Schematic representation of the hybrid charging model, where the quantum battery and the charger spins are mediated by a magnon mode of a YIG sphere.}\label{yig}
\end{figure}

As demonstrated in Fig.~\ref{yig}, the hybrid charging model for our quantum battery is consisted of a quantum charger $C$ (an $N$-spin system), a quantum battery $B$ (an $M$-spin system), and a magnon system (in the Kittel mode) acting as a mediator between $C$ and $B$. In the rotating-wave approximation, the total Hamiltonian can be written as ($\hbar\equiv1$)
\begin{equation}\label{H}
\begin{aligned}
&H=\omega_m m^\dag m+H_C+H_B+H_I, \\
&H_C=\sum_{i=1}^N\omega_{Ci}\sigma^+_{Ci}\sigma^-_{Ci}+\sum_{i\ne j}^N J_{ij}\left(\sigma^+_{Ci}\sigma^-_{Cj}+{\rm H.c.}\right), \\
&H_B=\sum_{k=1}^M\omega_{Bk}\sigma^+_{Bk}\sigma^-_{Bk}+\sum_{k\ne l}^M J_{kl}\left(\sigma^+_{Bk}\sigma^-_{Bl}+{\rm H.c.}\right), \\
&H_I=\sum_{i=1}^Ng_{i}\sigma^+_{Ci} m+\sum_{k=1}^M g_{k}\sigma^+_{Bk}m+{\rm H.c.}.
\end{aligned}
\end{equation}
The first term of $H$ is the magnon Hamiltonian, with $m$ ($m^\dag$) the annihilation (creation) operator of the magnon mode in the YIG sphere and $\omega_m$ its frequency~\cite{magnon,magnonhybrid}. The second and the third terms are the charger and the battery Hamiltonians, respectively. By convention, the spins in both quantum battery and charger are set to be resonant, i.e., $\omega_{Ci}=\omega_{Bk}=\omega$. $\sigma^+_j$ and $\sigma^-_j$ are respectively the Pauli raising and lowering operators for the $j$th spins. $J_{ij}$ or $J_{kl}$ is the direct coupling strength between battery or charger spins. The last term $H_I$ describes the mediated interaction between the magnon mode and the two spin systems, with $g_{i}$ or $g_{k}$ the particular coupling-strength. The interaction Hamiltonian can be tuned on/off by manipulating $\omega_m$ to be near-resonant/far-off-resonant from the spin frequency, which is applicable to the stable charging process.

In the large-detuning (dispersive) regime, $g_{i}$, $g_{k}$, $J_{ij}$, $J_{kl}\ll\omega_m$, $\omega$, $|\omega-\omega_m|$, the effective Hamiltonian can be obtained with the second-order perturbation theory~\cite{perturbation},
\begin{equation}\label{Heffss}
\begin{aligned}
H_{\rm eff}&=\sum_{i=1}^N \sum_{k=1}^M G_{ik}\left(\sigma^+_{Ci}\sigma^-_{Bk}+\sigma^-_{Ci}\sigma^+_{Bk}\right)\\
&+\sum_{i\ne j}^N (G_{ij}+J_{ij})\left(\sigma^+_{Ci}\sigma^-_{Cj}+\sigma^-_{Ci}\sigma^+_{Cj}\right)\\
&+\sum_{k\ne l}^M (G_{kl}+J_{kl})\left(\sigma^+_{Bk}\sigma^-_{Bl}+\sigma^-_{Bk}\sigma^+_{Bl}\right),
\end{aligned}
\end{equation}
where $G_{ik}=g_{i}g_{k}/(\omega-\omega_m)$. The derivation detail can be found in appendix~\ref{appa}. In the language of adiabatic elimination~\cite{adiabatic}, the far-off-resonant magnon mode can be eliminated to the second order of the weak coupling strengths, giving rise to the indirect couplings described by $G_{ik}$ (charger-battery), $G_{ij}$ (charger-charger), and $G_{kl}$ (battery-battery).

We emphasize again that in our model of Eq.~(\ref{H}) there is no direct coupling between the charger and the battery. By virtue of Eq.~(\ref{Heffss}), one can study a quantum battery system charged through a quantum wire played by the magnon mode. And the quantum wire (magnon) remains at its initial state all the time.

We denote by $\rho(t)$, $\rho_C(t)={\rm Tr}_B[\rho(t)]$, and $\rho_B(t)={\rm Tr}_C[\rho(t)]$ the density matrices for the whole system, the charger system, and the battery system, respectively. We can therefore identify the charger energy with $E_C(t)\equiv{\rm Tr}[H_C\rho_C(t)]$ and the battery energy with $E_B(t)\equiv{\rm Tr}[H_B\rho_B(t)]$. Initially, it is assumed that the charger is prepared in a full-charged state while the battery and the magnon are in their ground states,
\begin{equation}\label{rho0}
\rho(0)=\rho_C(0)\otimes|0\rangle_B\langle 0|\otimes|0\rangle_m\langle 0|,
\end{equation}
such that $E_C(0)>0$ and $E_B(0)=0$. This choice follows the energy-preserving protocols in which all the energy stored in the quantum battery $B$ at the end of the charging process that originates, without any thermodynamical ambiguity, from the charger $C$. The performance of the charger-battery setup can thus be studied in terms of the energy stored in the battery and the corresponding average charging power, defined respectively as,
\begin{equation}\label{ep}
\begin{aligned}
&E(t)\equiv E_B(t)={\rm Tr}[H_B\rho_B(t)],\\
&P(t)\equiv E(t)/t.
\end{aligned}
\end{equation}
The period for $E(t)$ attaining the maximum possible value $E_{\rm max}$ is set as the charging time $\tau$. And $P_{\rm max}$ is defined as the maximum value during the period of $t\in[0,\tau]$, which is usually larger than $P(\tau)=E(\tau)/\tau$ and can be obtained through an optimization evaluation.

\section{Magnon-mediated charging process}\label{process}

Here we focus on the time evolution by which the energy transferred from the charger to the quantum battery. To simplify the discussion but with no loss of generality, we set $g_{i}=g_C$, $g_{k}=g_B$ and $J_{ij}=J_{kl}=J$, i.e., the fluctuations around the average interaction strength are omitted. Then the effective Hamiltonian in Eq.~(\ref{Heffss}) can be rewritten as
\begin{equation}\label{Heff}
\begin{aligned}
H_{\rm eff}&=\sum_{i=1}^N \sum_{k=1}^M G\left(\sigma^+_{Ci}\sigma^-_{Bk}+\sigma^-_{Ci}\sigma^+_{Bk}\right)\\
&+\sum_{i\ne j}^N (G+J)\left(\sigma^+_{Ci}\sigma^-_{Cj}+\sigma^-_{Ci}\sigma^+_{Cj}\right)\\
&+\sum_{k\ne l}^M (G+J)\left(\sigma^+_{Bk}\sigma^-_{Bl}+\sigma^-_{Bk}\sigma^+_{Bl}\right),
\end{aligned}
\end{equation}
where $G=-g_Cg_B/\Delta$ with $\Delta\equiv\omega_m-\omega$ is the magnon-induced coupling strength between the charger and the battery spins. When the mode frequency of the magnon system $\omega_m$ is larger than the spin frequency $\omega$, $\Delta$ is positive as the convention in this work. Consequently, $G$ is found to be negative as required in the following optimized energy-transfer between the charger and the battery.

The coupling strength between magnon and spin is about $g/2\pi\approx1$ MHz~\cite{spinmagnoncoupling}. If the detuning frequency is taken as $\Delta=10g$, then the effective coupling strength $|G|/2\pi\approx 0.1$ MHz. In the configuration of one charger spin charging one battery spin (see the ensuing section), the charging time for our long-range quantum battery is about $50\mu$s, that will be definitely reduced wtih more charger spins (see Sec.~\ref{multione}).

\subsection{Energy transfer from one spin to one spin}\label{oneone}

\begin{figure}[htbp]
\centering
\includegraphics[width=0.4\textwidth]{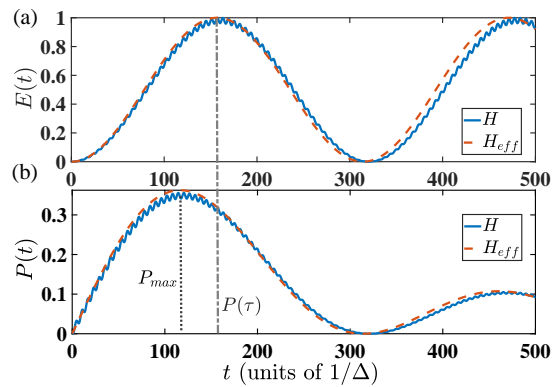}
\caption{(Color online) The dynamics of (a) $E(t)$ (in unit of $\omega$) and (b) $P(t)$ (in unit of $|G|\omega$) as functions of the dimensionless time in the ``one (charger spin) - to - one (battery spin)'' configuration. Here we fixed $g_C=g_B=0.1\Delta$ (in this case the effective coupling strength $G/\Delta=-0.01$). The initial state for the charger-battery system is taken as $|e\rangle_C\otimes|g\rangle_B$.}\label{spinmagnon}
\end{figure}

We begin by studying the simplest yet a nontrivial scenario, where both charger and battery are consisted of a single spin, i.e., $N=1$ and $M=1$. In this case, the effective Hamiltonian in Eq.~(\ref{Heff}) is reduced to $H_{\rm eff}=G(\sigma^+_C\sigma^-_B+\sigma^-_C\sigma^+_B)$, which is naturally independent on $J$. Considering the initial state $|e\rangle_C\otimes|g\rangle_B$, the evolved state of the charger-battery system can be expressed as
\begin{equation}\label{varphi1}
|\varphi_1(t)\rangle=\cos(|G|t)|e\rangle_C|g\rangle_B-i\sin(|G|t)|g\rangle_C|e\rangle_B,
\end{equation}
which yields the quantities in Eq.~(\ref{ep}),
\begin{equation}\label{EP1}
E(t)=\omega\sin^2(|G|t), \quad  P(t)=\omega\sin^2(|G|t)/t.
\end{equation}
Hence the maximum energy that can be stored in the quantum battery is $E_{\rm max}=\omega$ and it is firstly achieved at the moment $\tau=\pi/(2|G|)$. Then the corresponding average charging power is $P(\tau)=2|G|\omega/\pi$.

The dynamics of both battery energy $E(t)$ and charging power $P(t)$ driven by the total Hamiltonian in Eq.~(\ref{H}) or the effective Hamiltonian in Eq.~(\ref{Heff}) are demonstrated in Fig.~\ref{spinmagnon}(a) and (b), respectively. Comparing the results from different Hamiltonians (the blue solid lines for the total Hamiltonian and the red dashed lines for the effective Hamiltonian), one can observe that the effective dynamics smoothes the negligible oscillations along the exact dynamics and follows the same profile. Also, it is found that the numerical simulations match with the analytical results in Eq.~(\ref{EP1}): the maximum energy $E_{\rm max}=E(\tau)$ is about $\omega$ and the maximum charging power $P_{\rm max}$ is about $0.35|G|\omega$ (see the distinguished points).

\subsection{Energy transfer from multi-spins to one spin}\label{multione}

\begin{figure}[htbp]
\centering
\includegraphics[width=0.4\textwidth]{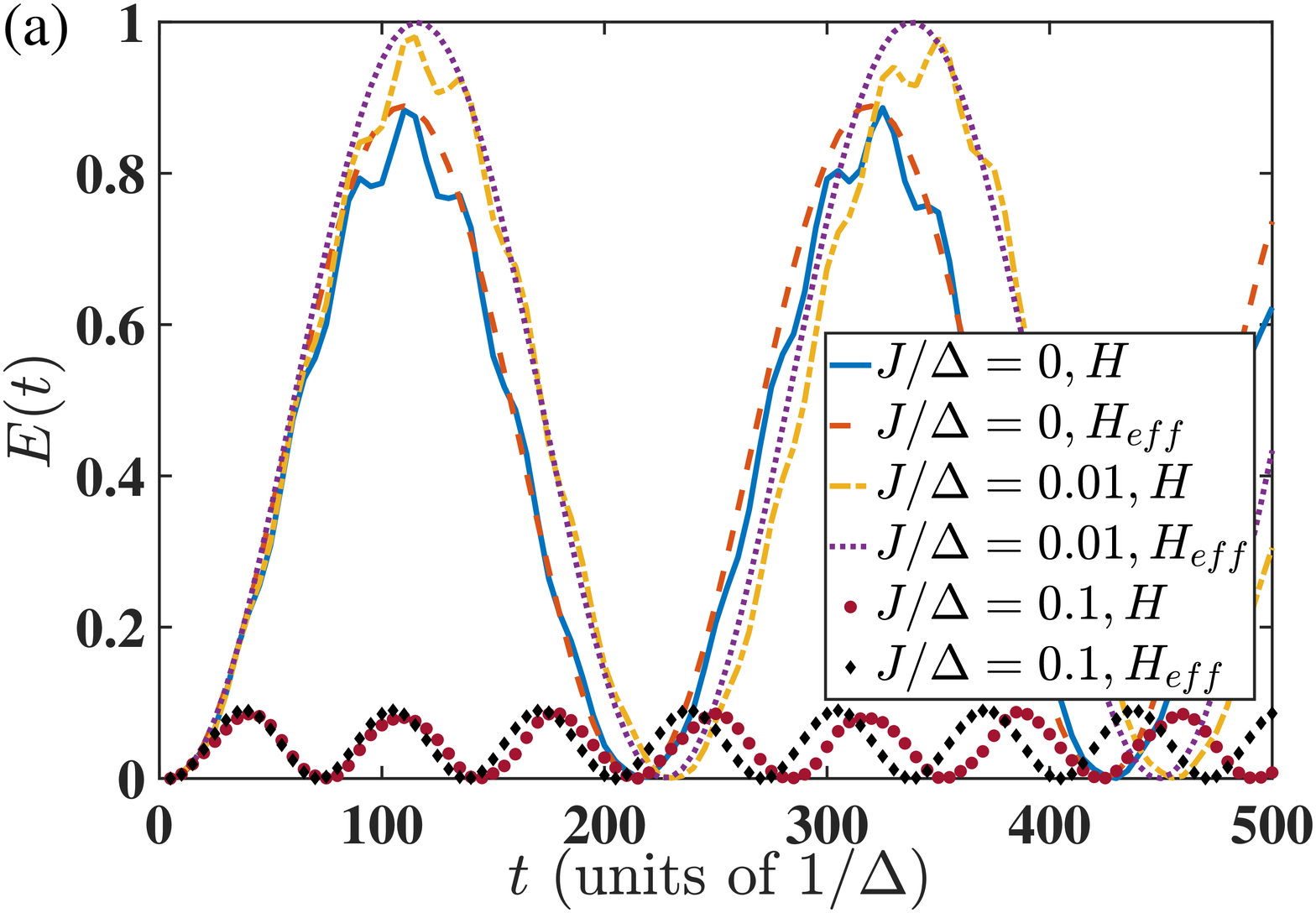}
\includegraphics[width=0.4\textwidth]{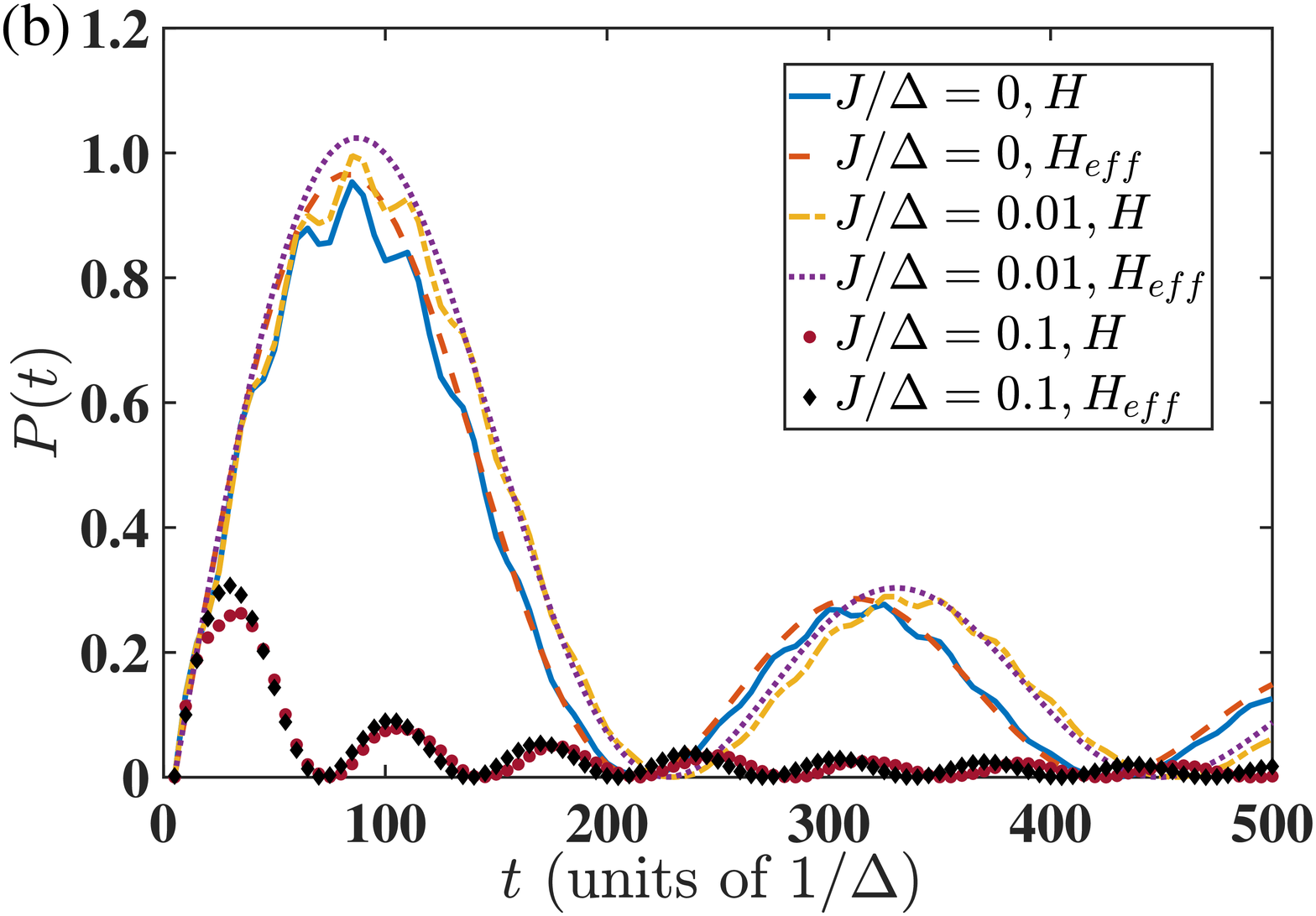}
\caption{(Color online) The dynamics of (a) $E(t)$ (in unit of $\omega$) and (b) $P(t)$ (in unit of $|G|\omega$) as functions of the dimensionless time in the ``two-to-one'' configuration. $g_C=g_B=0.1\Delta$ and then $G/\Delta=-0.01$. The initial state is $|ee\rangle_C|g\rangle_B$.}\label{spinmagnon2}
\end{figure}

In this section, we consider the configuration of $N>1$ and $M=1$. More spins in the charger could improve the average charging power, i.e., reduce the charging time for the battery to be full-charged. To see the effect of more charger spins, we first take $N=2$ and $M=1$ as an example. Initially, the two spins of the charger $C$ are in their excited state and the spin of the battery $B$ is in its ground state. Due to the fact that the exciton number of the whole system is conserved, the effective Hamiltonian in Eq.~(\ref{Heff}) can be written in the subspace spanned by $\{|eeg\rangle\equiv|e\rangle_{C1}|e\rangle_{C2}|g\rangle_B,|ege\rangle,|gee\rangle\}$:
\begin{equation}\label{Heff2}
H_{\rm eff}=\begin{bmatrix}
0 & G & G\\
G & 0 & G+J\\
G & G+J & 0\\
\end{bmatrix}.\\
\end{equation}
Then from the initial state $|eeg\rangle$, the time-evolved state of the charger-battery system can be written as
\begin{equation}\label{varphi2s}
\begin{aligned}
|\varphi_2(t)\rangle&=\left(e^{-i\epsilon_+t}-e^{-i\epsilon_-t}\right)
\frac{\sin(2\theta)}{2\sqrt{2}}(|ege\rangle+|gee\rangle)\\
&+\left(e^{-i\epsilon_+t}\sin^2\theta+e^{-i\epsilon_-t}\cos^2\theta\right)|eeg\rangle,
\end{aligned}
\end{equation}
where $\theta\equiv\tan^{-1}[2\sqrt{2}G/(G+J)]/2$ and $\epsilon_{\pm}=[G+J\pm\sqrt{(G+J)^2+8G^2}]/2$. The stored energy of the quantum battery in this state is,
\begin{equation}\label{EP2s}
E(t)=\omega\sin^2(2\theta)\frac{1-\cos(\epsilon_+t-\epsilon_-t)}{2}.
\end{equation}
We find that the stored energy can achieve its maximum value $E_{\rm max}=\omega\sin^2(2\theta)$ when $\cos(\epsilon_+t-\epsilon_-t)=-1$. That could be optimized at the sweet spot with $\sin^2(2\theta)=1$, that gives rise to $G+J=0$. Note $G=-g_Cg_B/\Delta=g_Cg_B/(\omega-\omega_m)$ is negative when $\omega_m>\omega$. The optimization condition $G+J=0$ could then be satisfied by tuning either the induced coupling strength of spins $G$ that is inversely proportional to the detuning $\Delta=\omega_m-\omega$ in magnitude or the inner coupling strength $J$ that is the dipole-dipole interaction between spins. The former can be manipulated by the external magnetic field and the latter depends on the mutual distances between spins. It is interesting to find when $J=0$, $\sin^2(2\theta)<1$, irrespective to the magnitude of $G$. Then $E_{\rm max}$ is always less than $\omega$, i.e., the battery can not be fully charged. In contrast, when $G+J=0$, the time evolution in Eq.~(\ref{varphi2s}) is reduced to
\begin{equation}\label{varphi2}
\begin{aligned}
|\varphi_2(t)\rangle&=\cos(\sqrt{2}|G|t)|eeg\rangle\\ &-i\frac{\sin(\sqrt{2}|G|t)}{\sqrt{2}}\left(|ege\rangle+|gee\rangle\right),
\end{aligned}
\end{equation}
yielding
\begin{equation}\label{EP2}
E(t)=\omega\sin^2(\sqrt{2}|G|t), \quad P(t)=\omega\sin^2(\sqrt{2}|G|t)/t.
\end{equation}
Hence the battery can be fully charged, i.e., $E_{\rm max}\equiv\omega$, at the moment $\tau=\pi/(2\sqrt{2}|G|)$. The corresponding average charging power is then $P(\tau)=2\sqrt{2}|G|\omega/\pi$, which is $\sqrt{2}$ times as that in the ``one-to-one'' configuration.

In Fig.~\ref{spinmagnon2}(a), we plot the dynamics of $E(t)$ under different inner coupling strengths $J$. For simplicity, we fix both the charger-magnon and the battery-magnon coupling strengths as $g_C=g_B=0.1\Delta$. Then $G/\Delta$ is fixed as $-0.01$, which provides the magnitude of the sweet spot for our numerical simulation. Comparing the lines without the inner coupling strength $J=0$ and with $J/\Delta=0.01$, one can observe that the maximum transferred energy attains $\omega$ when $J/\Delta=0.01=-G$, showing a clear advantage over the case under $J/\Delta=0$. The inner spin-spin coupling strength $J$ of both battery and charger does not always enhance the charging performance in comparison to the noninteracting case. It is found that the two dotted lines with $J/\Delta=0.1=-10G$ manifest a clear decline of $E(t)$ as well as $E_{\rm max}$ in comparison to the preceding results. In physics, when the direct coupling $J$ neutralizes the indirect coupling $G$, it can be used to enhance the charging performance by optimizing the maximum stored energy $E_{\rm max}$. However, if $J$ overwhelms $|G|$, then the free Hamiltonian of either charger or battery becomes dominant in the whole Hamiltonian as indicated by Eq.~(\ref{Heff}). A sufficiently large $J$ will effectively separate the dynamics of charger and battery and then suppress the energy-transfer process from charger to battery, which is determined by the interaction Hamiltonian. One can actually estimate the range of $J$ for showing the advantage in terms of the charged energy. With a fixed $G$, it is straightforward to find that $E_{\rm max}$ increases monotonically with $J$ when $0\leq J\leq-G$ and then decreases monotonically with $J$ when $J\geq-G$. The fact that $E_{\rm max}(J=-2G)=E_{\rm max}(J=0)$ implies that the charging performance in the noninteracting case will overwhelm that in the interacting case when $J>-2G$. The dynamics of the average charging power associated with Fig.~\ref{spinmagnon2}(a) is plotted in Fig.~\ref{spinmagnon2}(b). The results also indicate that at the optimized point for $J=-G$, one can have the strongest quantum battery.

In Figs.~\ref{spinmagnon2}(a) and (b), we provide the results from the original Hamiltonian $H$ and the effective Hamiltonian $H_{\rm eff}$ to justify our derivation. It is shown when $J\leq|G|=0.01\Delta$, the results obtained from different Hamiltonian almost match with each other. While when $J=0.1\Delta=10|G|$, the periods of both $E(t)$ and $P(t)$ under $H_{\rm eff}$ are less than those under $H$. Although $J$ is not greatly less than $\Delta$ in this case, the effective Hamiltonian up to the second order is still sufficient to capture the realistic dynamics in the first several periods of charging and discharging.

\begin{figure}[htbp]
\centering
\includegraphics[width=0.4\textwidth]{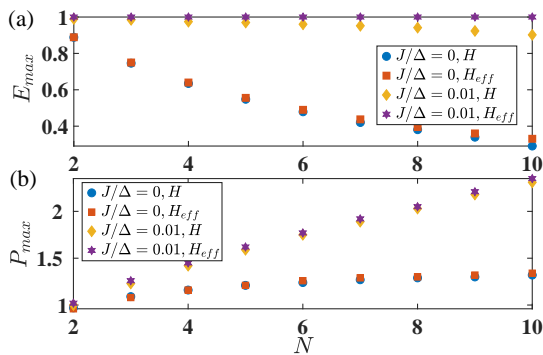}
\caption{(Color online) (a) The maximum energy $E_{\rm max}$ (in unit of $\omega$) and (b) the maximum average charging power $P_{\rm max}$ (in unit of $|G|\omega$) under either the total Hamiltonian in Eq.~(\ref{H}) or the effective Hamiltonian in Eq.~(\ref{heffs}) with various number of charger spins in the ``N-to-1'' configuration. $g_C=g_B=0.1\Delta$ and then $G/\Delta=-J/\Delta=-0.01$. The initial state is $|e\rangle^{\otimes N}_C|g\rangle_B$. }\label{spinmagnonN1}
\end{figure}

The optimization condition $G+J=0$ applies to more general scenarios. Now the charger is extended to $N$ spins and the battery remains as a single spin. The effective Hamiltonian in Eq.~(\ref{Heff}) can then be written as
\begin{equation}\label{heffs}
H_{\rm eff}=\left(\sum^{N}_{i=1}\sigma^+_{Ci}\right)\sigma^-_B+\left(\sum^{N}_{i=1}\sigma^-_{Ci}\right)\sigma^+_B.
\end{equation}
With the initial state $|e\rangle^{\otimes N}_C|g\rangle_B$ and the condition $J=-G$, the time evolution of the charger-battery system is
\begin{equation}\label{varphin}
\begin{aligned}
|\varphi_N(t)\rangle&=\cos(\sqrt{N}|G|t)|e\rangle^{\otimes N}|g\rangle \\
&-i\frac{\sin(\sqrt{N}|G|t)}{\sqrt{N}}\sum_{i=1}^N |e\rangle^{\otimes(i-1)}|g\rangle|e\rangle^{\otimes(N-i)}|e\rangle, \\
\end{aligned}
\end{equation}
where the last element in both items indicates the battery state. Then the stored energy of battery $E(t)$ and the average charging power $P(t)$ are respectively given by
\begin{equation}\label{EPN}
E(t)=\omega\sin^2(\sqrt{N}|G|t), \quad P(t)=\omega\sin^2(\sqrt{N}|G|t)/t.
\end{equation}
The maximum energy $E_{\rm max}=\omega$ can then be available and the charging time is $\tau=\pi/(2\sqrt{N}|G|)$. Consequently, the corresponding average charging power $P(\tau)=2\sqrt{N}|G|\omega/\pi$ scales with $\sqrt{N}$, demonstrating a collective advantage similar to the Dicke quantum battery~\cite{highpowerbattery}.

In Fig.~\ref{spinmagnonN1}, we present the stored energy of battery and its average charging power with an increasing number of charger-spin $N$. For the maximum energy $E_{\rm max}$ in Fig.~\ref{spinmagnonN1}(a), under the optimized situation $G+J=0$, the induced coupling cancels the inner coupling and then holds $E_{\rm max}$ close to its ideal value $\omega$. In the absence of $J$, $E_{\rm max}$ rapidly decreases by increasing $N$, i.e., more chargers will reduce rather than enhance the battery efficiency. For the maximum charging power $P_{\rm max}$ in Fig.~\ref{spinmagnonN1}(b), it is increased by increasing $N$ under the optimized situation, as expected by the preceding analytical result. It shows an advantage over the power under the vanishing inner coupling strength, that approaches a steady value when $N\geq6$. In addition, more charger spins will undermine the effective Hamiltonian. One can find that the realistic $E_{\rm max}$ will gradually depart from the ideal result with more charger spins by comparing the yellow diamonds and the purple hexagons in Fig.~\ref{spinmagnonN1}(a).

\subsection{Energy transfer from multi-spins to multi-spins}\label{multimulti}

At the sweet spot of our charging protocol $G+J=0$, the effective Hamiltonian in Eq.~(\ref{Heff}) for arbitrary $N$ and $M$ is reduced to
\begin{equation}\label{Heffn}
H_{\rm eff}=\sum_{i=1}^N\sum_{k=1}^MG\left(\sigma^+_{Ci}\sigma^-_{Bk}+\sigma^-_{Ci}\sigma^+_{Bk}\right).
\end{equation}
Either of the charger or the battery could be regarded as a collective non-interacting spins. By using the collective spin operators
\begin{equation}\label{J}
\hat{J}_{\pm}^C=\sum^N_{i=1}\sigma^{\pm}_{Ci}, \quad \hat{J}_{\pm}^B=\sum^M_{k=1}\sigma^{\pm}_{Bk},
\end{equation}
the effective Hamiltonian can be rewritten as
\begin{equation}\label{Heffh}
H_{\rm eff}=G\left(\hat{J}_-^C\hat{J}_+^B+\hat{J}_+^C\hat{J}_-^B\right).
\end{equation}

The collective spin operators $\hat{J}_{\pm}$ satisfy the commutation relation of the usual angular momentum. Under the eigen-basis of $\hat{J}_z\equiv\sum_i\sigma_i^z$, we have
\begin{equation}\label{jj}
\hat{J}_{\pm}|j,m\rangle=\sqrt{(j\pm m+1)(j\mp m)}|j,m\pm1\rangle,
\end{equation}
where $m$ is the eigenvalue of $\hat{J}_z$, an integer running from $-j$ to $j$. For the charger system, $j=N/2$ and $m=-N/2,-N/2+1,\cdots,N/2$, and for the battery system, $j=M/2$ and $m=-M/2,-M/2+1,\cdots,M/2$. Because the energy quantum number $j$ is constant during the evolution under $H_{\rm eff}$, it can be omitted in the following discussion.

\begin{figure}[htbp]
\centering
\includegraphics[width=0.4\textwidth]{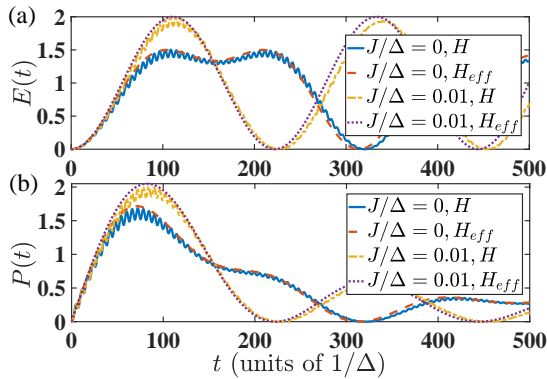}
\caption{(Color online) The ``two-to-two'' dynamics of (a) $E(t)$ (in unit of $\omega$) and (b) $P(t)$ (in unit of $|G|\omega$) as functions of the dimensionless time. $g_C=g_B=0.1\Delta$. The initial state is $|ee\rangle_C|gg\rangle_B$. }\label{spinmagnonbattery2}
\end{figure}

The optimization condition $G+J=0$ can be firstly justified in a special case, in which both the charger and the battery are consisted of two spins. The initial state is $|\varphi'_2(0)\rangle=|ee\rangle_C|gg\rangle_B$, which is equivalent to the state $|m=1\rangle_C|m=-1\rangle_B$ by virtue of the angular momentum operators in Eq.~(\ref{J}). Then the effective Hamiltonian in Eq.~(\ref{Heffh}) for this ``two-to-two'' configuration can be written in the subspace spanned by $\{|1,-1\rangle\equiv|1\rangle_C|-1\rangle_B,|0,0\rangle,|-1,1\rangle\}$, which reads,
\begin{equation}\label{heffhs}
H_{\rm eff}=2\begin{bmatrix}
0 & G  & 0\\
G & 0 & G\\
0 & G & 0\\
\end{bmatrix}.
\end{equation}
Subsequently, the time evolved state is
\begin{equation}\label{varphi22}
\begin{aligned}
|\varphi'_2(t)\rangle&=\cos^2\left(\sqrt{2}|G|t\right)|1,-1\rangle-\sin^2\left(\sqrt{2}|G|t\right)|-1,1\rangle\\
&+\frac{i}{2}\sin\left(2\sqrt{2}|G|t\right)|0,0\rangle,
\end{aligned}
\end{equation}
and then the stored energy is
\begin{equation}\label{Et2}
E(t)=\omega\sin^2\left(\sqrt{2}|G|t\right)\left[1+\sin^2\left(\sqrt{2}|G|t\right)\right].
\end{equation}
The battery thus gets fully charged with the maximum energy $2\omega$ when $\tau=\pi/(2\sqrt{2}|G|)$. Consequently, the average charging power reads $P(\tau)=4\sqrt{2}|G|\omega/\pi$, which is double of that in the ``two-to-one'' configuration.

\begin{figure}[htbp]
\centering
\includegraphics[width=0.4\textwidth]{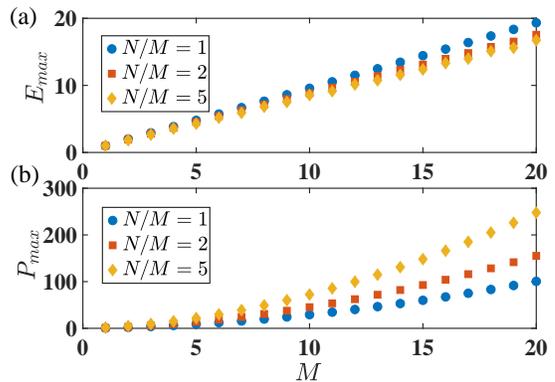}
\caption{(Color online) (a) The maximum energy $E_{\rm max}$ (in unit of $\omega$) and (b) the maximum average charging power $P_{\rm max}$ (in unit of $|G|\omega$) as functions of the number of battery spins $M$. Results of $N/M=1,2,5$, are plotted with the blue dots, the red squares and the yellow diamonds, respectively. $g_C=g_B=0.1\Delta$. The initial state is $|e\rangle^{\otimes N}_C|g\rangle^{\otimes M}_B$.}\label{spinmagnonHNM}
\end{figure}

The numerical simulation about the dynamics of $E(t)$ and $P(t)$ is demonstrated in Fig.~\ref{spinmagnonbattery2}. For this ``two-to-two'' configuration, our effective Hamiltonian in Eq.~(\ref{heffhs}) still captures the pattern of the dynamics by the total Hamiltonian in Eq.~(\ref{H}). At the sweet spot $G+J=0$, the dynamics of $E(t)$ in Fig.~\ref{spinmagnonbattery2}(a) is rectified to a sinusoidal pattern and the quantum battery is full-charged with energy $2\omega$. In contrast, the stored energy of the battery is at most $1.5\omega$ in the absence of the inner coupling, namely $J=0$ (see both the blue solid line and the red dashed line). Accordingly, the average charging power is significantly enhanced in Fig.~\ref{spinmagnonbattery2}(b).

To see the charging performance in the general ``$N$-to-$M$'' configuration, we present the maximum energy $E_{\rm max}$ and the corresponding maximum charging power $P_{\rm max}$ in Fig.~\ref{spinmagnonHNM}, where the results are organized according to various ratio of $N/M$. In Fig.~\ref{spinmagnonHNM}(a), $E_{\rm max}$ increases in a linear way with the size of the quantum battery when $N/M=1$. Under more charger spins with $N/M>1$, the linear scaling of $E_{\rm max}$ becomes sub-linear. So that more charger spins will decline the stored energy for each cell in a multipartite quantum battery. Yet in Fig.~\ref{spinmagnonHNM}(b), one can find that an opposite pattern manifests in the maximum power $P_{\rm max}$: a larger $N/M$ and more charger spins generate a stronger charging power.

\section{Charging process in the presence of systematic errors}\label{qsdprocess}

Our charging protocol relies on the quantum wire supported by the magnon mode, which is fully determined by the external magnetic field. Due to the fact that this mode is an effective model about an ensemble of magnetic spins through the Holstein-Primakoff (HP) transformation~\cite{magnon,supermode,supermode1}, a systematic error on the Zeeman splitting of spins might be induced by the spatial fluctuation of magnetic field in the YIG sphere. In this section, our attention is turned to the quantum wire. We employ a non-Markovian quantum-state-diffusion equation to provide an exact solution on the preceding stochastic errors in the magnon mode. As an example, the ``one-to-one'' configuration, i.e., one battery spin charged by one charger spin mediated by the magnon mode, is used to check the robustness of our protocol. The total Hamiltonian in this scenario can be written as
\begin{equation}\label{H2}
\begin{aligned}
H&=[\omega_m+\epsilon(t)]m^\dag m+\omega\left(\sigma^+_C\sigma^-_C+\sigma^+_B\sigma^-_B\right)\\
&+g\left(\sigma^+_C+\sigma^+_B\right)m+g\left(\sigma^-_C+\sigma^-_B\right)m^\dag,
\end{aligned}
\end{equation}
where $\epsilon(t)$ is a real Gaussian noise denoting a semi-classical dephasing channel. It is assumed that $M[\epsilon(t)]=0$ and $M[\epsilon(t)\epsilon(s)]=\alpha(t,s)$, where $M[\cdot]$ means an ensemble average and $\alpha(t,s)$ is the correlation function.

In the rotating frame with respect to $S=[\omega_m t+\mathcal{E}(t)]m^\dag m$, where $\mathcal{E}(t)\equiv \int^t_0 ds \epsilon(s)$, the full Hamiltonian in Eq.~(\ref{H2}) becomes
\begin{equation}\label{Hi}
\begin{aligned}
H'_I&=e^{iS}He^{-iS}-\dot{S} \\ &=\omega(\sigma^+_C\sigma^-_C+\sigma^+_B\sigma^-_B)+[ge^{-i\mathcal{E}(t)}(\sigma^+_C+\sigma^+_B)m+{\rm H.c.}].
\end{aligned}
\end{equation}
The QSD approach starts from the Schr\"{o}dinger equation
\begin{equation}\label{psit}
i\partial |\Psi(t)\rangle/\partial t=H'_I |\Psi(t)\rangle,
\end{equation}
where $|\Psi(t)\rangle$ represents the full wavefunction of spins and magnons. The occupation number of magnon is supposed to be vanishing, that is consistent with the prerequisite for the HP transformation. Then the initial overall wavefunction is $|\Psi(0)\rangle\approx|\psi_{t=0}\rangle|0\rangle_m$, where $|\psi_{t=0}\rangle$ is the battery-charger state at time $0$. We define that a stochastic wavefunction (trajectory) for the battery-charger system $|\psi_t(z^*)\rangle\equiv\langle z||\Psi(t)\rangle$, where $||z\rangle$ is a Bargmann coherent state for the magnon mode with a displacement amplitude $z$ normally distributed in the whole complex plane. After a standard derivation~\cite{qsd0,qsd,qsds,qsds2,qsd4,qsd5}, one can obtain the QSD equation from Eq.~(\ref{psit}):
\begin{equation}\label{hpsi}
\begin{aligned}
\partial_t|\psi_t(z^*)\rangle&=-iH_{\rm QSD}|\psi_t(z^*)\rangle, \\
&=\left[-iH_{sys}+Lz^*_t-L^\dag\bar{O}(t,z^*)\right]|\psi_t(z^*)\rangle
\end{aligned}
\end{equation}
where $H_{\rm sys}=\omega(\sigma^+_C\sigma^-_C+\sigma^+_B\sigma^-_B)$ and $L=\sigma^-_C+\sigma^-_B$ denote the battery-charger system Hamiltonian and the system-magnon coupling operator, respectively. The quantum trajectory $|\psi_t(z^*)\rangle$ is desired to recover the density operator of the system by taking ensemble average: $\rho_t=M[|\psi_t(z^*)\rangle\langle \psi_t(z^*)|]=\int\frac{dz^2}{\pi}e^{-|z|^2}|\psi_t(z^*)\rangle\langle\psi_t(z^*)|$.

The complex random process $z_t^*=-igz^*\exp[i\omega_mt-i\mathcal{E}(t)]$ combines a Jaynes-Cummings (JC) process [in the absence of the energy-fluctuation term, Eq.~(\ref{H2}) descrites a JC-type Hamiltonian] and a dephasing noise, which are indicated by $\exp(i\omega_mt)$ and $\exp[-i\mathcal{E}(t)]$, respectively. These two processes are statistically independent with each other due to their irrelevant sources. The correlation function for $z_t$ is then found to be
\begin{equation}\label{gt}
\begin{aligned}
&G(t,s)=M[z_t z^*_s]=g^2 e^{-i\omega_m(t-s)}M\left[e^{-i\mathcal{E}(t)+i\mathcal{E}(s)}\right]\\
&=g^2 e^{-i\omega_m(t-s)}\exp\left[\int^t_sdt_1\int^{t_1}_sdt_2\alpha(t_1,t_2)\right].
\end{aligned}
\end{equation}
For simplicity, the dephasing noise is described by the well-known Ornstein-Uhlenbek noise. Specifically, $\alpha(t,s)=\frac{\Gamma\gamma}{2}e^{-\gamma|t-s|}$, where $\Gamma$ is the coupling strength between the magnon mode and the magnetic environment and $1/\gamma$ characterizes the memory capacity of the dephasing source. The correlation function in Eq.~(\ref{gt}) thus turns out to be
\begin{equation}\label{gts}
\begin{aligned}
G(t-s)&=g^2e^{-i\omega_m(t-s)}\times \\ & \exp\left\{-\frac{\Gamma}{2}\left[(t-s)+\frac{e^{-\gamma(t-s)}-1}{\gamma}\right]\right\}.
\end{aligned}
\end{equation}

\begin{figure}[htbp]
\centering
\includegraphics[width=0.4\textwidth]{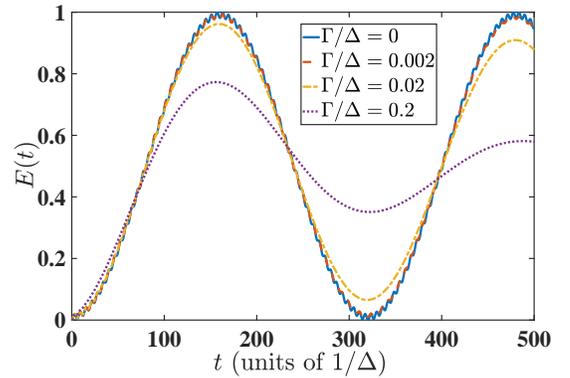}
\caption{(Color online) Dynamics of $E(t)$ (in unit of $\omega$) as a function of a dimensionless time under the systematic error, which is induced by the fluctuation in the magnon-mode frequency with various noise strength $\Gamma$. Here we fixed $g_C=g_B=0.1\Delta$. The initial state for the charger-battery system is $|e\rangle_C\otimes|g\rangle_B$. }\label{qsdgamma}
\end{figure}

The system operator $\bar{O}(t,z^*)$ of $H_{\rm QSD}$ in Eq.~(\ref{hpsi}) is defined by $\bar{O}(t,z^*)\equiv\int^t_0dsG(t,s)O(t,s,z^*)$. The O-operator $O(t,s,z^*)$ in our system can be exactly expressed by
\begin{equation}\label{ot}
\begin{aligned}
O(t,s,z^*)&=f_1(t,s)O_1+f_2(t,s)O_2\\
&+i\int^t_0 ds' f_3(t,s,s')z^*_{s'}O_3,
\end{aligned}
\end{equation}
where $O_1=\sigma^-_C+\sigma^-_B$, $O_2=\sigma^-_C\sigma^z_B+\sigma^z_C\sigma^-_B$, $O_3=\sigma^-_C\sigma^-_B$, and the coefficient functions of $f_j$, $j=1,2,3$, are determined by the boundary condition $O(t,t,z^*)=L$ and the consistency condition~\cite{qsd1,qsd3}. In the absence of the double-exciton state $|e\rangle_C|e\rangle_B$, that is the case for our charging protocol, the O-operator $O(t,s,z^*)$ is reduced to $O(t,s)=f_1(t,s)O_1+f_2(t,s)O_2$ and then $\bar{O}(t,z^*)=\bar{O}(t)=F_1(t)O_1+F_2(t)O_2$, where $F_j(t)\equiv\int^t_0dsG(t,s)f_j(t,s)$, $j=1,2$. Under the Markovian approximation, i.e., $\gamma\rightarrow\infty$, it is found that
\begin{equation}\label{Ft}
\begin{aligned}
&\dot{F}_1=g^2+(i\omega-i\omega_m-\Gamma/2)F_1+F^2_1+3F^2_2,\\
&\dot{F}_2=(i\omega-i\omega_m-\Gamma/2)F_2-F^2_1+F^2_2+4F_1F_2.\\
\end{aligned}
\end{equation}
And the boundary conditions are given by $F_1(0)=F_2(0)=0$. In the subspace spanned by $\{|eg\rangle=|e\rangle_C|g\rangle_B,|ge\rangle,|gg\rangle\}$, the stochastic Hamiltonian $H_{\rm QSD}$ in Eq.~(\ref{hpsi}) can be written as
\begin{equation}\label{Heffs}
H_{\rm QSD}=\begin{bmatrix}
\omega-i\mathcal{F} & -i\mathcal{F} & 0\\
-i\mathcal{F} & \omega-i\mathcal{F} & 0\\
iz^*_t &iz^*_t &0\\
\end{bmatrix}.
\end{equation}
Here the parameter $\mathcal{F}(t)\equiv F_1(t)-F_2(t)$, which satisfies
\begin{equation}\label{ft}
\dot{\mathcal{F}}=g^2+(-i\Delta-\Gamma/2)\mathcal{F}+2\mathcal{F}^2
\end{equation}
according to Eq.~(\ref{Ft}) and $\mathcal{F}(0)=0$. Considering the initial state for the charger-battery system $|eg\rangle$, its time evolution is described by~\cite{qsd4}
\begin{equation}\label{phit}
|\psi_t(z^*)\rangle=a(t)|eg\rangle+b(t)|ge\rangle+c(t)|gg\rangle,
\end{equation}
where $a(t)=\exp[-2\int^t_0ds\mathcal{F}(s)]/2+1/2$ and $b(t)=\exp[-2\int^t_0ds\mathcal{F}(s)]/2-1/2$. Eventually the stored energy in the quantum battery is found to be
\begin{equation}\label{Et}
E(t)=|b(t)|^2\omega.
\end{equation}

In Fig.~\ref{qsdgamma}, we plot the dynamics of stored energy $E(t)$ with various noise magnitude $\Gamma/\Delta=0,0.002,0.02,0.2$. One can find that all these dynamics effectively simulate an energy relaxation process, which is the result of the combination of a dephasing noise from the frequency-fluctuation of the magnon mode and a dissipation-less JC interaction between the spins and the magnon mode. It is also indicated by the correlation function in Eq.~(\ref{gts}) under $\gamma\rightarrow\infty$, which describes an Ornstein-Uhlenbek dissipation noise in regard to the coupling operator $L=\sigma^-_C+\sigma^-_B$. So that one can see gradual decreasing oscillating-magnitudes in all these lines. Under a strong noise with $\Gamma/\Delta=0.2$ (see the purple dotted line), the battery could keep nearly $0.8\omega$ at most in the stored energy during the first quasi-period. When $\Gamma/\Delta=0.02$ (see the yellow dot-dashed line), $E_{\rm max}\approx0.96\omega$ during the first quasi-period. And under an even weaker noise with $\Gamma/\Delta=0.002$ (see the red dashed line), $E_{\rm max}$ is almost the same as the ideal result. Therefore our charging protocol is robust to the systematic errors.

\section{Discussion and conclusion}\label{conclusion}

Our mediated-charging protocol for quantum battery can be experimentally established in a hybrid system consisted of a magnon mode and two groups of many-spin emitters, such as the diamond nitrogen-vacancy defects or the silicon-vacancy defects. The neutralization of the indirect coupling mediated by the magnon mode and the direct coupling of spins $G+J=0$ is demanded to optimize the charging performance of our protocol. In particular, the indirect coupling strength $G=g^2/\Delta$ with $\Delta\equiv\omega-\omega_m$ could be modulated by tuning the frequency $\omega_m$ of the magnon mode and the single magnon-spin coupling strength $g$. $\omega_m$ depends on the external magnetic field. $g\sim2\pi\times1$ MHz can enter the strong-coupling regime~\cite{spinmagnoncoupling,spinmagnoncouplings} when the radius of YIG sphere is $30$ nm and the distance between YIG and spins is $36$ nm. In this case, the corresponding effective coupling strength $|G|$ is about $10$ KHz to $100$ KHz when the detuning $|\Delta|$ is tuned from $100g$ to $10g$. On the other hand, the magnitude of $J$ is found to be dozens of KHz~\cite{direct,directcoupling}. Then the frequency-match condition $G=-J$ can be realized through the precise control over the magnetic field.

An analog protocol could be implemented in similar platforms, such as the circuit-QED systems~\cite{cqed,cqedultrastrong,cqedphysics}, where the quantum wire is provided by the microwave cavity-mode instead of the magnon mode and the superconducting qubits play the role of the charger and battery spins. Note the coupling strength between the mediator and the charger and battery cells can reach the ultrastrong regime ($g/\omega\ge 0.1$ and $g/2\pi \sim 10^2\sim10^3$ MHz) in recent experiments~\cite{cqedultrastrong}. Consequently, the charging speed can be accelerated by two or three orders in magnitude in comparison to the current spin-magnon system at the cost of an extremely low-temperature setup.

In conclusion, we have presented a long-range quantum charging protocol in a hybrid spin-magnon system. In our protocol, both battery and charger are decoupled many-interacting-spins, and they are coupled to the Kittle mode of the magnon system. The magnon mode is regarded as a ``charging wire" connecting the charger and battery, which allows to realize a stable charging protocol due to the fact that it can be switched on and off by tuning the magnon frequency via adjusting the bias magnetic field. We find that the direct spin-spin interactions among charger and battery can be used to significantly promote the charging performance nearby an optimized point $J=-G$. We analysis the stored energy and charging power under various sizes of both charger and battery. We also discuss the decoherence raised by the stochastic fluctuation of the magnon frequency with the QSD equation and justify the robustness of our protocol against this systematic error. Our work provides a novel implementation of the quantum battery with discrete-variable chargers and extends the application range of the spin-magnon system.

\section*{Acknowledgments}

We acknowledge grant support from the National Science Foundation of China (Grants No. 11974311 and No. U1801661), and Zhejiang Provincial Natural Science Foundation of China under Grant No. LD18A040001.

\appendix

\section{The effective Hamiltonian}\label{appa}

This appendix is contributed to obtain the effective Hamiltonian in Eq.~(\ref{Heffss}), that describes the energy transfer from charger to battery mediated by a quantum wire (the magnon system) as shown in Fig.~\ref{yig}. One can apply the standard perturbation theory to the original Hamiltonian in Eq.~(\ref{H}) with respect to the coupling strengths $g_{i}, g_{k}\ll\omega,\omega_m,\Delta$, with $\Delta\equiv\omega-\omega_m$. Here the spins in both battery and charger are resonant, i.e., $\omega_{Ci}=\omega_{Bk}=\omega$.

It is instructive to first focus on the energy transfer between a single spin-$i$ and another spin-$k$ mediated by the magnon. The relevant charging process occurs in the subspace spanned by $\{|eng\rangle\equiv |e\rangle_{i}|n\rangle_m|g\rangle_{k},|gne\rangle\}$, in which the subscript implies the state for spin or magnon mode. The system Hamiltonian can be written as
\begin{equation}\label{Hsub}
\begin{aligned}
H&=H_0+H_I,\\
H_0&=\omega_m m^\dag m+\omega\left(\sigma^+_{i}\sigma^-_{i}+\sigma^+_{k}\sigma^-_{k}\right),\\
H_I&=g_{i}\sigma^+_{i}m+g_{k}\sigma^+_{k} m+{\rm H.c.}.
\end{aligned}
\end{equation}
To the second order of the perturbation Hamiltonian $H_I$, the effective coupling strength between any pair of eigenstates $|p\rangle$ and $|q\rangle$ of the unperturbed Hamiltonian $H_0$ in Eq.~(\ref{Hsub}) is given by~\cite{perturbation,perturbation2}
\begin{equation}\label{secondp}
G=\sum_{w\neq p,q}\frac{\langle q|H_I|w\rangle\langle w|H_I|p\rangle}{E_p-E_w},
\end{equation}
where $E_w$ is the eigenenergy associated with the eigenstate $|w\rangle$ of $H_0$, namely, $E_w=\langle w|H_0|w\rangle$. Note the condition $\omega_m\neq\omega$ lifts the degeneracy in $H_0$.

We consider the contributions to the effective coupling strength from the two paths connecting $|eng\rangle$ and $|gne\rangle$, i.e., $|eng\rangle \to |g(n+1)g\rangle \to |gne\rangle$ and $|eng\rangle \to |e(n-1)e\rangle \to |gne\rangle$ due to $H_I$. Then by virtue of Eq.~(\ref{secondp}), we have
\begin{equation}\label{Geff}
G_{ik}=\frac{g_{i}g_{k}}{\omega-\omega_m}.
\end{equation}
The full Hamiltonian in Eq.~(\ref{Hsub}) thus could be approximated by a second-order effective Hamiltonian:
\begin{equation}\label{Heffsub}
H_{\rm eff}=G_{ik}\left(\sigma^+_{i}\sigma^-_{k}+\sigma^-_{i}\sigma^+_{k}\right).
\end{equation}

The preceding derivation as well as the formation of the effective Hamiltonian applies to all the indirect spin-spin interaction mediated by the common magnon mode. The effective Hamiltonian describing the overall Hamiltonian in Eq.~(\ref{H}) can therefore be written as
\begin{equation}\label{Happendix}
\begin{aligned}
H_{\rm eff}&=\sum_{i=1}^N \sum_{k=1}^M G_{ik}(\sigma^+_{Ci}\sigma^-_{Bk}+{\rm H.c.})\\
&+\sum_{i\ne j}^N (G_{ij}+J_{ij})(\sigma^+_{Ci}\sigma^-_{Cj}+{\rm H.c.})\\
&+\sum_{k\ne l}^M (G_{kl}+J_{kl})(\sigma^+_{Bk}\sigma^-_{Bl}+{\rm H.c.}).
\end{aligned}
\end{equation}

\bibliographystyle{apsrevlong}
\bibliography{reference}

\end{document}